\documentclass[preprint,pre,amsmath,amssymb,showpacs]{revtex4}

\usepackage{graphicx}
\usepackage{dcolumn}
\usepackage{bm}
\usepackage{verbatim}
\usepackage{subfigure}
\usepackage{dcolumn}

\newtheorem{conjecture}{Conjecture}
\newcommand{\myemph}{}

\begin{document}

\title{Packing dimers on $(2p + 1) \times (2q + 1) $  lattices}

\author{Yong Kong}
\email{matky@nus.edu.sg}
\affiliation{%
Department of Molecular
Biophysics and Biochemistry\\
W.M. Keck Foundation Biotechnology Resource Laboratory \\
Yale University\\
333 Cedar Street, New Haven, CT 06510\\
email: \texttt{yong.kong@yale.edu} 
}%

\date{\today}

\begin{abstract}
We use computational method to 
investigate the number of ways to pack dimers on \emph{odd-by-odd} lattices.
In this case, 
there is always a single vacancy in the lattices.
We show that the 
dimer configuration numbers on $(2k+1) \times (2k+1)$ \emph{odd} square 
lattices have some remarkable number-theoretical properties
in parallel to those of close-packed dimers
on $2k \times 2k$ \emph{even} square lattices, 
for which exact solution exists. 
Furthermore, we demonstrate that there is an unambiguous logarithm term
in the finite size correction of free energy of odd-by-odd lattice strips
with any width $n \ge 1$.
This logarithm term determines the distinct behavior of the
free energy of odd square lattices.
These findings reveal a deep and previously 
unexplored connection between statistical physics models
and number theory, and indicate the possibility that
the monomer-dimer problem might be solvable.

%
\end{abstract}

\pacs{05.50.+q, 02.10.De, 02.70.-c}


\maketitle

\section{Introduction}

The dimer model has been investigated for many years 
to represent adsorption of diatomic molecules on a surface.
In the model the surface is represented as 
regular plane lattice and the diatomic molecules as rigid dimers, 
which fit into the lattice 
so that each dimer occupies two adjacent lattice sites and no lattice site 
is covered by more than one dimer.  
The model is closely related to another well-studied 
lattice model, the Ising model. 
A central problem of the model is to enumerate the 
dimer configurations on the lattice.  
Exact solution exists only for the special case 
when the lattice of size $m \times n$ is completely covered by dimers 
if at least one of $m$ and $n$ is even 
(the close-packed dimer problem) \cite{Kasteleyn1961,Temperley1961}. 
The general monomer-dimer problem where there are vacancies in the lattice, 
like the general Ising model in an 
external field, however, remains unsolved and is usually considered 
computationally intractable \cite{Jerrum1987}. 
One exception is a recent analytic solution to the 
special case where there is a single 
vacancy at certain specific sites on the boundary 
of the lattice \cite{Tzeng2003}.
  
In this report we use computational method to 
investigate the number of ways to pack dimers on 
odd-by-odd  lattices.
In this case, 
there is always a single vacancy in the lattices.
The computation method is outlined in Section \ref{S:comp}.
In Section \ref{S:square}
we show that the 
leading coefficients of the partition functions ($a_N$) on odd 
$(2k+1) \times (2k+1)$ square lattices 
have some surprising number-theoretical properties, 
in parallel to those found in even $2k \times 2k$ square 
lattices \cite{John2000}.  
We also show that  
the average free energy of odd lattices of finite sizes 
differs dramatically from that of even lattices. 
Unlike that of even lattices,
the average free energy of odd lattices
approaches the thermodynamic limit non-monotonically,
descending first for small lattices
to reach a minimum before ascending as the size of lattice increases.
In contrast,
the free energy of even lattices  
approaches the thermodynamic limit monotonically.
The behavior of the free energy of finite size odd lattices
also differs from that of lattices where there is a single 
vacancy restricted at certain specific sites on the boundary 
of the lattices \cite{Tzeng2003}.
In this latter case,
the free energy behaves in much the same way as that of
close-packed even lattices without vacancy.

To investigate further the origin of the different behaviors
of free energy in odd and even lattice, in Section \ref{S:FSC}
we calculate
the free energy on strip lattices of size $m \times n$,
where both $m$ and $n$ are \emph{odd}.
The data show unambiguously that there is a logarithm term
in the free energy of odd lattices, which is absent in the
free energy when $m$ or $n$ is even.
We demonstrate that it is this logarithm term in the finite size correction
of the  free energy in odd lattices that leads to the
non-monotonicity and minimum in the free energy of the odd square
lattices.
In Section \ref{S:discussion} we discuss our findings in a broader 
prospective, and explore their potential 
relation with the theory of computational complexity.


\section{Computational method} \label{S:comp}

Here we use computational method to enumerate the dimers on plane lattices.
The computational method we use 
is similar to the one introduced previously where symmetry 
of the dimer configurations is used to obtain a recurrence of the 
partition functions and to reduce the complexity 
of the problem \cite{Kong1999}.
The computational method is outlined 
below.  For a lattice of size $m \times n$, 
the property we are interested in is the configurational grand 
canonical partition function 
\[
 Z_{m,n}(x, y, z) = \sum_{N_x, N_y, N_z} 
 g_{m,n}(N_x, N_y, N_z) x^{N_x} y^{N_y} z^{N_z} 
\]
where $x$, $y$, and $z$ are the activities of the $x$-dimers, $y$-dimers,
and monomers, respectively,
and $g_{m,n}(N_x, N_y, N_z)$ is the number of ways to place
$N_x$ dimers in the $x$ (horizontal) direction, 
$N_y$ dimers in the $y$ (vertical) direction, 
and $N_z$ monomers on the sites that are not occupied by dimers.
Without losing generality, we can let $z=1$, so that the partition function
can be written as
\begin{equation} \label{E:Z}
 Z_{m,n}(x, y) = \sum_{N_x, N_y} 
 g_{m,n}(N_x, N_y) x^{N_x} y^{N_y}.
\end{equation}

Since there is no interaction between dimers except for the
constraint that no site can be occupied by more than one dimer,
the partition function of a strip lattice with width of $n$
is totally determine by the 
configurations of dimers on two adjacent rows of the lattice, 
each row itself being 
a one-dimensional linear lattice of size $n$ \cite{Kong1999}. 
A square matrix $M_n$ is set up based on these two rows (see below).
The vector $\Omega_m$, which consists of partition function of Eq. \ref{E:Z}
as well as other \emph{contracted} partition functions \cite{Kong1999} 
on a  $m \times n$ strip, is calculated by the 
following recurrence
\begin{equation} \label{E:rec}
 \Omega_m = M_n \Omega_{m-1}.
\end{equation}                                                       

\begin{figure}
  \centering
  \includegraphics[width=\columnwidth]{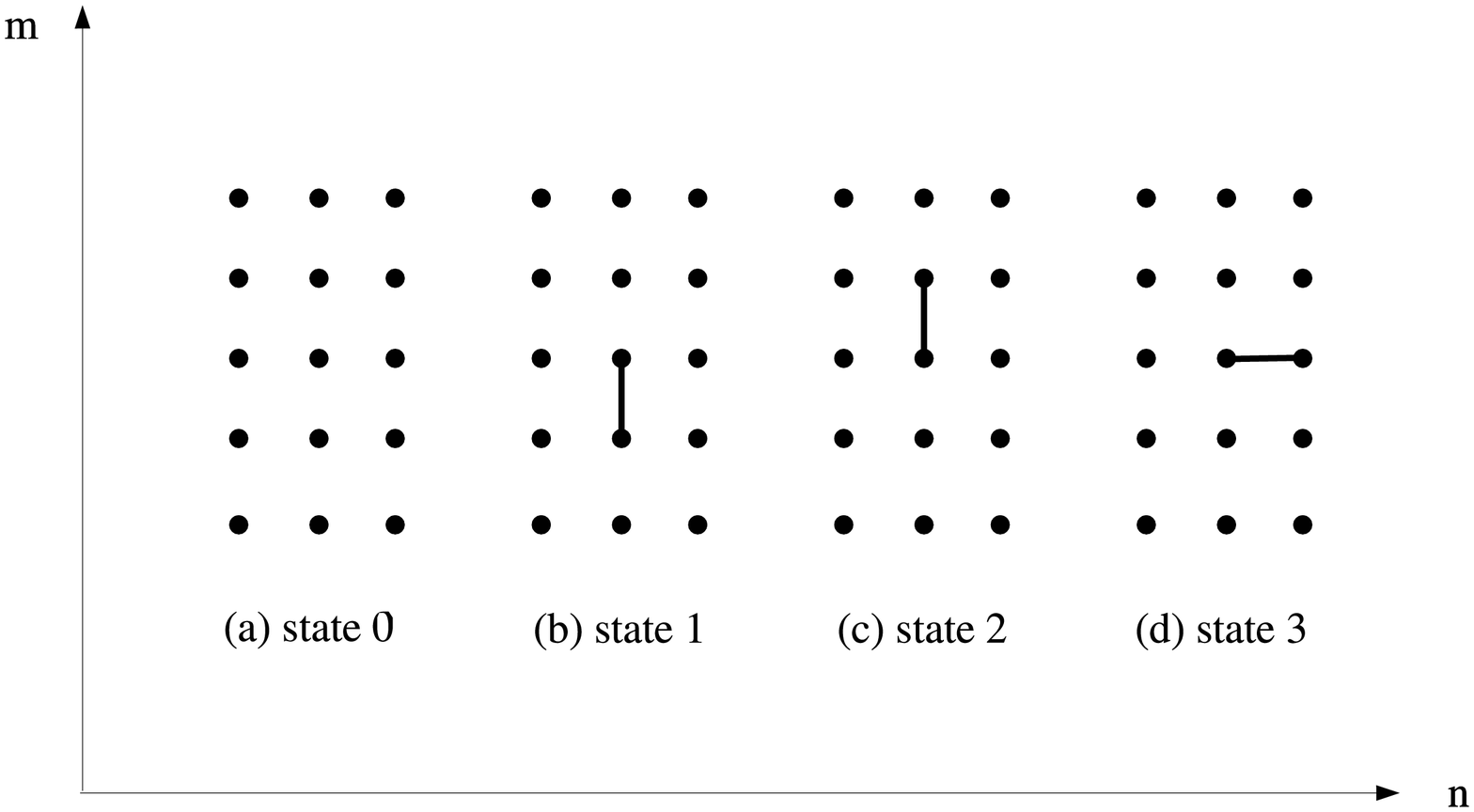}
  \caption{The configurational states of one lattice site.
    Suppose the strip with width $n$ is expanding in the vertical
    direction.  The four states that each site can have are depicted 
    here for the \emph{central site} in the figure. 
    (a) State $0$: the site is empty;
    (b) State $1$: the site is occupied by the first half of a vertical dimer;
    (c) State $2$: the site is occupied by the second half of 
    a vertical dimer;
    (d) State $3$: the site is occupied by a horizontal dimer.
    Since the lattice strip is growing vertically,
    for horizontal dimers we do not need to distinguish
    first half or second half.
    \label{F:states}} 
\end{figure}

To construct the matrix $M_n$, we notice that each lattice site
can have four dimer configurational states, as shown in Figure~\ref{F:states}
for the center site:
(a) State $0$: the monomer state, where the site is empty;
(b) State $1$: the site is occupied by the first half of a vertical dimer;
(c) State $2$: the site is occupied by the second half of 
a vertical dimer;
(d) State $3$: the site is occupied by a horizontal dimer.
Since the lattice strip is growing vertically 
($n$ is fixed and $m$ is changing),
for horizontal dimers we do not need to distinguish
first half or second half.
For each lattice site $(i, j)$, we denote $s(i, j)$ as its state.

The total number of dimer configurations $t(n)$
in a one-dimensional lattice with length $n$ is $t(n) = 4^n$.
Some of these configurations are
not valid: when dimers occupy horizontally, they have to occupy even
number of consecutive sites.  
The total number of valid configurations $v(n)$ is
given by the generating function
$1/(1-3x-x^2)$. This generating function can be derived from
the obvious recurrence $v(n) = 3 v(n-1) + v(n-2)$.

The size and elements of matrix $M_n$ are determined by the 
unique relative configurations of the dimers on two adjacent rows
of the lattice.
The basic method of Ref. \onlinecite{Kong1999} is to express 
\emph{contracted} partition functions of one row with 
contracted partition functions of the next row.
A particular contracted partition function is
the partition function of the lattice
when states of lattice sites in one row are fixed in a
given combinations. 
For dimers on row $j+1$,  
there are only two \emph{effective} 
configurational states of dimers on row $j$
if the states of dimers in row $j$ are fixed
in a given configuration.
When dimers on row $j$ are in states $s(i, j) = 0$, $1$, or $3$, 
they will contribute the same expansion coefficients
when the contracted partition functions of row $j$ are expressed 
with the contracted partition functions of row $j+1$.
In other words, for the purpose of building up the matrix $M_n$ to
calculate the partition functions recursively in the vertical direction,
the occupancy of a horizontal dimer on site $(i, j)$ has the same
effect as having an empty site on site $(i, j)$.
So does the occupancy of the first half of a dimer on site $(i, j)$. 
By grouping these states together, and taking symmetry into account,
the total number of \emph{unique} configurational states 
is given by formula
$
  u(n) = 2^{n-1} + 2^{\lfloor (n+1)/2\rfloor - 1}
$.

The detailed algorithm to construct matrix $M_n$ is given below.
(1) For a strip lattice with a given width of $n$,
the $t(n) = 4^n$ configurations are enumerated.
(2) Only $v(n)$ valid configurations are kept.
Those that are not valid are filtered out.
(3) Based on symmetry and the grouping of dimer states
mentioned above, each valid
configuration is assigned into a group.
(4) For each of the $u(n)$ groups, say group $p$, 
pick a configuration $c_p(\alpha)$ in the group,
and loop through all the valid configurations determined in step (2)
to check their compatibility with $c_p(\alpha)$.

The possible \emph{compatible} combinations of 
states in lattice sites $(i, j)$ and $(i, j+1)$
are $\{ ( s(i, j) \in \{0, 1, 3\}$ and $s(i, j+1) \in \{0, 1, 2\} )$
or  $( s(i, j) = 2$ and $s(i, j+1) = 1 )\}$.
All other combinations of states are not feasible.
They either lead to contradiction to the definition of the states,
or violate
the constraint that each lattice site cannot be multiply occupied.
For example, the combination $\{ s(i, j) = 0$ and $s(i, j+1) = 1 \}$
belongs to the first category, while $\{ s(i, j) = 2$ and $s(i, j+1) = 2 \}$
belongs to the second category.

Suppose the configuration being checked for compatibility with $c_p(\alpha)$
is from group $q$, and is labeled as $c_q(\beta)$.
The check of compatibility is carried out for all the lattice sites
along the horizontal direction for the configuration,
that is, for $i = 1, \dots, n$. Only when all sites of $c_q(\beta)$
are compatible with sites of $c_p(\alpha)$
does the configuration $c_q(\beta)$ make contributions to the matrix element
$M_n(p, q)$. 
As an example, the matrix for $n=3$ is given below.
\[
\mathit{M_3}= \left[ \begin {array}{cccccc} 1+2\,x&2\,y+2\,xy&y&2\,{y}^{2}
&{y}^{2}&{y}^{3}\\\noalign{\medskip}1+x&y&y&{y}^{2}&0&0
\\\noalign{\medskip}1&2\,y&0&0&{y}^{2}&0\\\noalign{\medskip}1&y&0&0&0&0
\\\noalign{\medskip}1&0&y&0&0&0\\\noalign{\medskip}1&0&0&0&0&0
\end {array} \right] 
\]

The following example gives the various sizes of $t(n)$, $v(n)$, and $u(n)$. 
For $n=19$,
the numbers of total, total valid, 
and total unique configurations are $t(19) = 274877906944$, 
$v(19) = 6616217487$, and $u(19) = 262656$.
Since the size of matrix $M_n$ is determined by $u(n)$,
for $n=19$, $M_{19}$ is of size $262656 \times 262656$,
with each element as a polynomial of $x$ and $y$.
All computations are carried out in a $32$-bit 
Linux workstation with 2G memory.
Exact integers, instead of floating number approximations,
are used in the calculations.
           
\section{Number-theoretical properties and free energy of packing dimers
on odd-by-odd square lattices} \label{S:square}

Once the matrix $M_n$ is set up, 
the partition functions can be obtained recursively
from Eq. \ref{E:rec}.
Although the full partition function of Eq. \ref{E:Z} can be obtained 
from the method, 
here we are especially interested in $a_N$
\footnote{ 
The coefficient of the leading term in the partition 
function gives directly the total work required to saturate the lattice 
with the $N$ dimers \cite{Wyman1964,diCera1995}:
\[
 \Delta G_N = -k_B T \ln a_N
\]  
where $k_B$ is the Boltzmann constant and $T$ is the temperature. 
The average work to ligate one dimer is given by $-(\ln a_N) / N$
in unit of $k_B T$. 
This concept is not very often 
mentioned in statistical physics literature,
but it plays an important role in studying 
cooperative binding phenomena in 
biology and chemistry \cite{Wyman1964,diCera1995}. 
The value of $x_m$ in $\ln x_m = -(\ln a_N) / N$ is usually called 
``Wyman median ligand activity'' \cite{Wyman1964} or 
``mean ligand activity'' \cite{diCera1995}.
}, 
the coefficient of the leading term in the partition 
function Eq. \ref{E:Z} when $x=y$,
where $N = \lfloor mn/2 \rfloor$:
\[
Z_{m,n}(x) = a_N x^N + a_{N-1} x^{N-1} + \cdots + a_0
\]

\begin{figure}
  \centering
  \includegraphics[angle=270,width=\columnwidth]{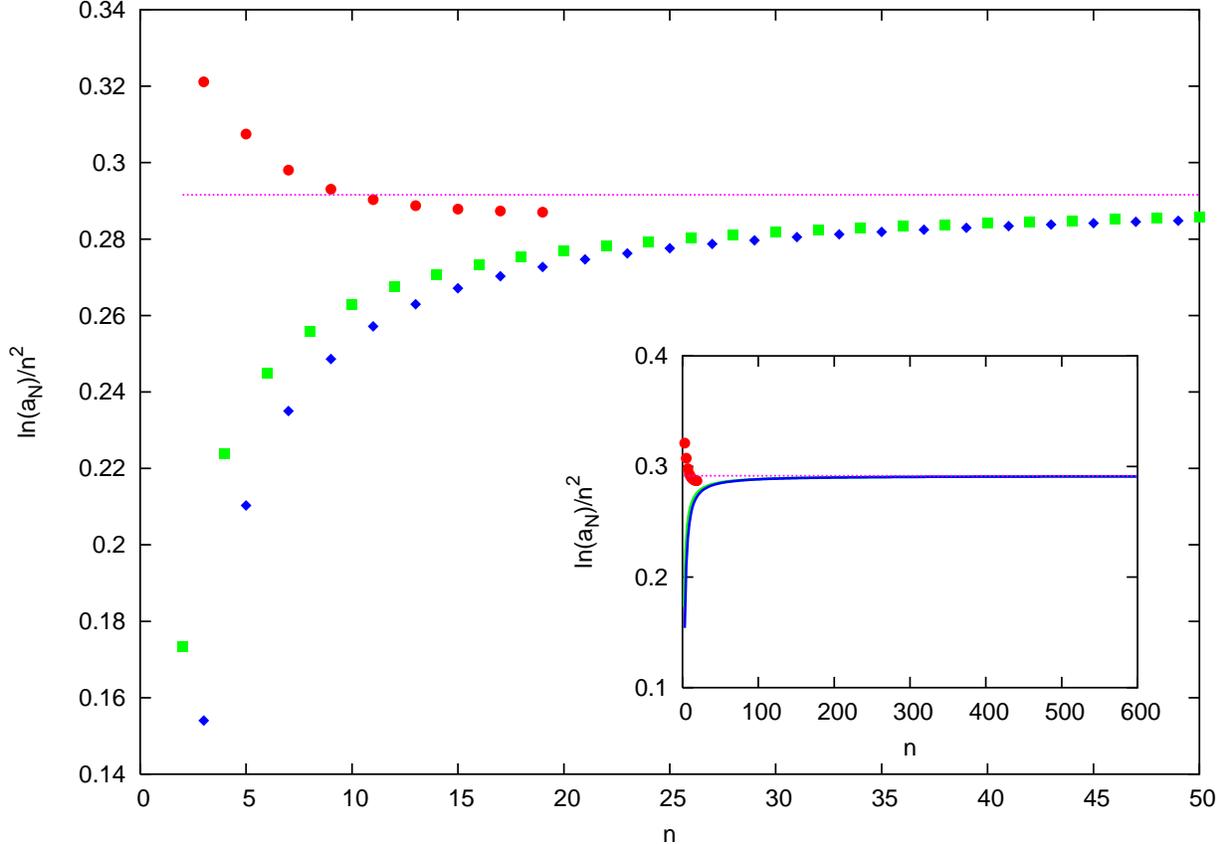}
  \caption{\label{F:a_N} 
(Color online) Free energy per lattice site ($\ln a_N/n^2$) 
of odd and even 
square lattices in unit of $-k_B T$ as a function of $n$,
the size of the lattice.
The values for \emph{odd} lattices are in solid circle, those for
\emph{even} lattices in solid square.  
The value in the thermodynamic
limit, $\lim_{n \rightarrow \infty} n^{-2} \ln a_N(n) = 0.291560904$, 
is also shown as the
dotted horizontal line. The values for even lattices beyond $n=20$
are calculated from the exact results \cite{Kasteleyn1961,Temperley1961}.
In the inset more data are shown for the even lattices to make it
clearer  the trend for $\ln a_N/n^2$ to approach the thermodynamic
limit.
Also shown in solid diamond 
are the values from lattices where there is a single
vacancy restricted at certain specific sites on the boundary of
the lattices \cite{Tzeng2003}. 
}
\end{figure}

When $n=2k$, the square lattice $n \times n$
can be fully covered by  $N=n^2/2$ dimers, 
and we can take advantage of the available exact results for this 
cases \cite{Kasteleyn1961,Temperley1961}. 
We refer this 
kind of lattices as \emph{even} lattices in the following. When $n=2k+1$, 
the square lattice 
 can at most accommodate $N=(n^2-1)/2$ dimers. In this case there is always a 
single vacancy in the lattice. We refer the lattices as \emph{odd} lattices. 
The free energy per lattice site 
$\ln a_N/ (mn)$ (divided by $-k_B T$) for the odd and even square lattices 
as a function of $n$, the size of the lattice, is shown 
in Figure~\ref{F:a_N}. 
The exact result for the even lattices \cite{Kasteleyn1961,Temperley1961}
shows that in the thermodynamic limit,
\[
 \lim_{n \rightarrow \infty} n^{-2} \ln a_N(n) = G/ \pi = 0.291560904
\]
where $G$ is the Catalan number.
For the fully-packed lattices, the difference between the 
even and odd lattices is the existence of a single vacancy site in the odd 
lattices. The effects 
of this single vacancy site are evident for small lattices, 
but as the size of the lattice 
increases, the effects become smaller. 
The parity of even and odd lattices will manifest its 
greatest effects in terms of $\ln a_N / n^2$  when $n$ is small. 
This is shown in Figure~\ref{F:a_N}. 
As can be seen from the figure, in the beginning for small lattices, 
the free energy of even and odd lattices approaches the 
thermodynamic limit from 
different directions, with that of even lattices 
smaller than 
the asymptotic value, 
and that of odd lattices greater than the asymptotic value. 
However, it is interesting to note that the 
free energy of odd lattices soon reaches and passes over the 
thermodynamic limit value at about $n=11$. 
Since $\ln a_N / n^2$ is a well-defined 
thermodynamic property, 
it should approach the same value 
in the thermodynamic limit 
as $n$ approaches infinity for both even and odd lattices. 
Accordingly it is expected that the free energy of
the odd lattices will reach a minimum, after which it would approach 
the thermodynamic 
limit monotonically above the values of the even lattices. 
The same conclusion can also 
be obtained from the existence theory of the 
dimer system, which asserts that the limit of 
$\ln a_N / n^2$  exists as $n$ approaches infinity \cite{Hammersley1966}. 
The behavior of the free energy of 
odd lattices is to contrast with that of the even lattices, 
which approaches the 
thermodynamic limit monotonically from the very beginning. 
The current calculation 
shows that the minimum of the odd lattices will be reached when  $n\ge19$.
We will discuss the origin of the minimum in the free energy of odd lattices
after we discuss the logarithm correction term
found in the finite size correction in Section~\ref{S:FSC}.

Also shown in Figure~\ref{F:a_N}
is the free energy of lattices where there is a single
vacancy restricted at certain specific sites on the boundary of
the lattices \cite{Tzeng2003}. The free energy for these lattices
approaches the 
thermodynamic limit monotonically, in the same way and the same direction
as that of even lattices, 
and behaves quite differently from that of odd lattices.
The data show that 
removing the restriction of the location of the vacancy 
on the lattice boundary
leads to
significant changes in the nature of the problem.

The values of $a_N$ up to $n=19$ are listed in the second column of 
Table \ref{T:a_N}. 
One of the advantages of using exact integer calculations 
instead of approximation calculations is that 
the number-theoretical nature of the coefficients of the partition functions 
can be studied in more details. 
For the even lattices $2k \times 2k$, 
it is first conjectured and later proved \cite{John2000} by using the 
explicit formulas \cite{Kasteleyn1961,Temperley1961} 
that $a_N = 2^k c_k = 2^k b_k^2$, where $c_k$ and $b_k$ are integers, and
$b_k$ has the following property:
\begin{equation*}
 b_k = \begin{cases}
   k+1 \pmod{2^4}              & \text{if $k$ is even,}\\ 
   (-1)^{(k-1)/2} \pmod{2^4}   & \text{if $k$ is odd.}
   \end{cases}
\end{equation*}
From these results it is 
easy to show that for even lattices: 
\begin{equation*}
c_k = b_k^2 = \begin{cases}
1  \pmod{2^4}   & \text{if $k=8p+q, q=0,1,6,7$,} \\ 
-7 \pmod{2^4}   & \text{if $k=8p+q, q=2,3,4,5$.} 
   \end{cases}
\end{equation*}
So for the $2k \times 2k$ lattices, 
the sequence of 
$a_N / 2^k \pmod{2^4}$  is, starting with $k=1$: 
$1$, $-7$, $-7$, $-7$, $-7$, 
$1$, $1$, $1$, $1$, $-7, \dots$.

\begingroup
\squeezetable
\begin{table*}
\caption{\label{T:a_N}The values of $a_N$,
the coefficients of the leading term in the partition 
functions of dimers on  $n \times n$ square lattices,
and their factorizations.}
\begin{ruledtabular}
\begin{tabular}{rllr}
 $n$  & $a_N$ & factors of $a_N = 2^k c_k$ & $c_k \mod 2^4$\\ \hline
 $2$  & $2$                            & $2$             &  $1$\\
 $\myemph{3}$  & $\myemph{18}$                           & $\myemph{2 \cdot 3^2}$   &  $\myemph{-7}$\\
 $4$  & $36$                           & $2^2 \cdot 3^2$ &  $-7$\\
 $\myemph{5}$  & $\myemph{2180}$                         & $\myemph{2^2 \cdot 5 \cdot 109}$ &  $\myemph{1}$\\
 $6$  & $6728$                         & $2^3 \cdot 29^2$ &  $-7$\\
 $\myemph{7}$  & $\myemph{2200776}$                      & $\myemph{2^3 \cdot 3 \cdot 107 \cdot 857}$ 
           &  $\myemph{-7}$\\
 $8$  & $12988816$                     & $2^4 \cdot 17^2 \cdot 53^2$ &  $-7$\\
 $\myemph{9}$  & $\myemph{20355006224}$                  & $\myemph{2^4 \cdot 7 \cdot 31 \cdot 5862617}$
           &  $\myemph{1}$\\
 $10$ & $258584046368$                 & $2^5 \cdot 241^2 \cdot  373^2$
           &  $-7$\\ 
 $\myemph{11}$ & $\myemph{1801272981919008}$             & $\myemph{2^5 \cdot 3 \cdot 13 \cdot 19 \cdot 139 \cdot 546508031}$ &  $\myemph{-7}$ \\
 $12$ & $53060477521960000$            & $2^6 \cdot 5^4 \cdot 31^2 \cdot 53^2 \cdot 701^2$ &  $1$\\
 $\myemph{13}$ & $\myemph{1560858753560238398528}$       & $\myemph{2^6 \cdot 313 \cdot 439 \cdot 177490360930511}$   &  $\myemph{1}$\\
 $14$ & $112202208776036178000000$     & $2^7 \cdot 3^{10} \cdot 5^6 \cdot 19^2 \cdot 29^4 \cdot 61^2$  &  $1$\\
 $\myemph{15}$ & $\myemph{13428038397958481723104394368}$ & $\myemph{2^7 \cdot 948517 \cdot 110600600710425473093}$  &  $\myemph{-7}$\\
 $16$ & $2444888770250892795802079170816$ & $2^8 \cdot 101^2 \cdot 9929^2 \cdot 5849^2 \cdot 16661^2$ &  $1$\\
 $\myemph{17}$ & $\myemph{1157111379933346772804754279450353920}$ & $\myemph{2^8 \cdot 5 \cdot 903993265572927166253714280820589} $ &  $\myemph{1}$\\
$18$  & $548943583215388338077567813208427340288$ & $2^9 \cdot 37^4 \cdot 457^2  \cdot 1597^2 \cdot 33629^2 \cdot 30817^2$ &  $1$\\
$\myemph{19}$  &  $\myemph{1004777133003025735713513459537724394989392384}$ & $\myemph{2^9 \cdot 3 \cdot 206870980007 \cdot 2578470886559 \cdot 1226356459698559963} $ &  $\myemph{-7}$\\
\end{tabular}
\end{ruledtabular}
\end{table*}
\endgroup

     From Table \ref{T:a_N} we can see that parallel but distinct properties 
also hold for the odd 
lattices. We have the following conjectures for odd lattice with size 
 $(2k+1) \times (2k+1)$:

\begin{conjecture} For odd lattice $(2k+1) \times (2k+1)$,  
$a_N$ can be factored as
\[
a_N = 2^k c_k
\]
where $c_k$ is an odd integer.  
Furthermore, when $k > 1$,
$c_k$ is squarefree: its prime decomposition contains no repeated factors. 
\end{conjecture}

\begin{conjecture} For odd lattice $(2k+1) \times (2k+1)$,  we have
\begin{equation*}
c_k = \begin{cases}
1  \pmod{2^4}      & \text{if $k$ is even,}\\
-7 \pmod{2^4}      & \text{if $k$ is odd.}  
\end{cases} 
\end{equation*}
The sequence of $a_N/2^k \pmod{2^4}$ is, 
starting from $k=1$: $-7$, $1$, $-7$, $1$, 
    $-7$, $1, \dots$. 
\end{conjecture}

Since very large numbers and big matrices are involved in the calculations,
it is important to check the correctness of the results.
The correctness of the data can be confirmed in several ways.
Firstly, 
the method mentioned in Section \ref{S:comp} applies to 
strip lattices with width $n$
no matter whether $n$ is even or odd.  
For even $n$, the exact results \cite{Kasteleyn1961,Temperley1961}
can be used to check the correctness of the results.
Secondly,
for a lattice strip of size $m \times n$ with a 
given width of $n$ when $n$ is odd, 
the recurrence of Eq.~\ref{E:rec}
will calculate \emph{both} even and odd values of $m$ using the \emph{same}
matrix $M_n$.
For those even values of $m$, exact formula \cite{Kasteleyn1961,Temperley1961}
can be used to check
the correctness of the results.  
For example, when $n=19$,
the values of $a_N$ calculated from Eq.~\ref{E:rec} 
for $m=18$ and $m=20$ are
$
83575338913384268749982379454805917136051
$
and
$
3911222696776012972239561518359338259546415835
$
respectively,
which are exactly what the exact formula gives.
The value $a_N$ of $m \times n  = 19 \times 19$ lattice is 
calculated from that of $18 \times 19$ lattice,
and the value $a_N$ of $20 \times 19$ lattice is 
calculated from that of $19 \times 19$ lattice.

Thirdly, the numbers can be cross-checked based on the fact that
$a_N$ for a lattice of size $m \times n$ should be the same
as that for a lattice of size $n \times m$, even though different matrices
$M_n$ and $M_m$ are involved.
For example, the value of $a_N$ for $19 \times 17$ lattice,
which is calculated from matrix $M_{17}$ of a size $65792 \times 65792$,
gives the same value as that calculated from $17 \times 19$ lattice,
which is calculated from matrix $M_{19}$.
The value is $18985861771720893262968550933152139575482$.
Much larger values have been checked for these tests,
some of which are shown in the following section.

\section{Finite size correction} \label{S:FSC}
The data shown in Figure~\ref{F:a_N} clearly indicate that
the finite size correction of free energy of 
odd lattices differs significantly from that of
even lattices. 
Finite size correction is investigated in \cite{Fisher1961,Ferdinand1967}
for the close-packed dimer model on $m \times n$ lattices
with at least one of $m$ or $n$ being even,
and in \cite{Tzeng2003}
for lattices where there is a single
vacancy restricted at certain specific sites on the boundary of the
lattices.
In these models, exact solutions are known
and finite size corrections can be obtained
from expansions of the exact expressions.

To investigate the finite size correction of free energy of
dimers in odd-by-odd lattices where there is always a vacancy,
we calculate the partition functions for long strip lattices $m \times n$
using Eq. \ref{E:rec} for $n=1, \dots, 19$,
and fit $a_N$ with the following function
\begin{equation} \label{E:fit}
 \frac{\ln a_N}{m n} = a_0 + \frac{a_1}{m} + \frac{a_2}{m^2} + 
 \frac{a_3}{m^3} + \frac{a_4}{m^4} + \frac{h}{n} \frac{\ln(m+1)}{m}.
\end{equation}
The reason to choose
$\ln(m+1)$ instead of $\ln(m)$ in Eq. \ref{E:fit} is that
for $n=1$, we have the following exact results:
\begin{equation} \label{E:n1}
 a_N = \begin{cases}
   (m+1)/2  & \text{$m$ is odd}\\
   1        & \text{$m$ is even}
   \end{cases}.  
\end{equation}
The fitting results are equally unambiguous if $\ln(m)$ were used
(data not shown).

\begin{figure}
    \begin{minipage}[b]{\columnwidth}
      \centering 
      \includegraphics[angle=270,width=\columnwidth]{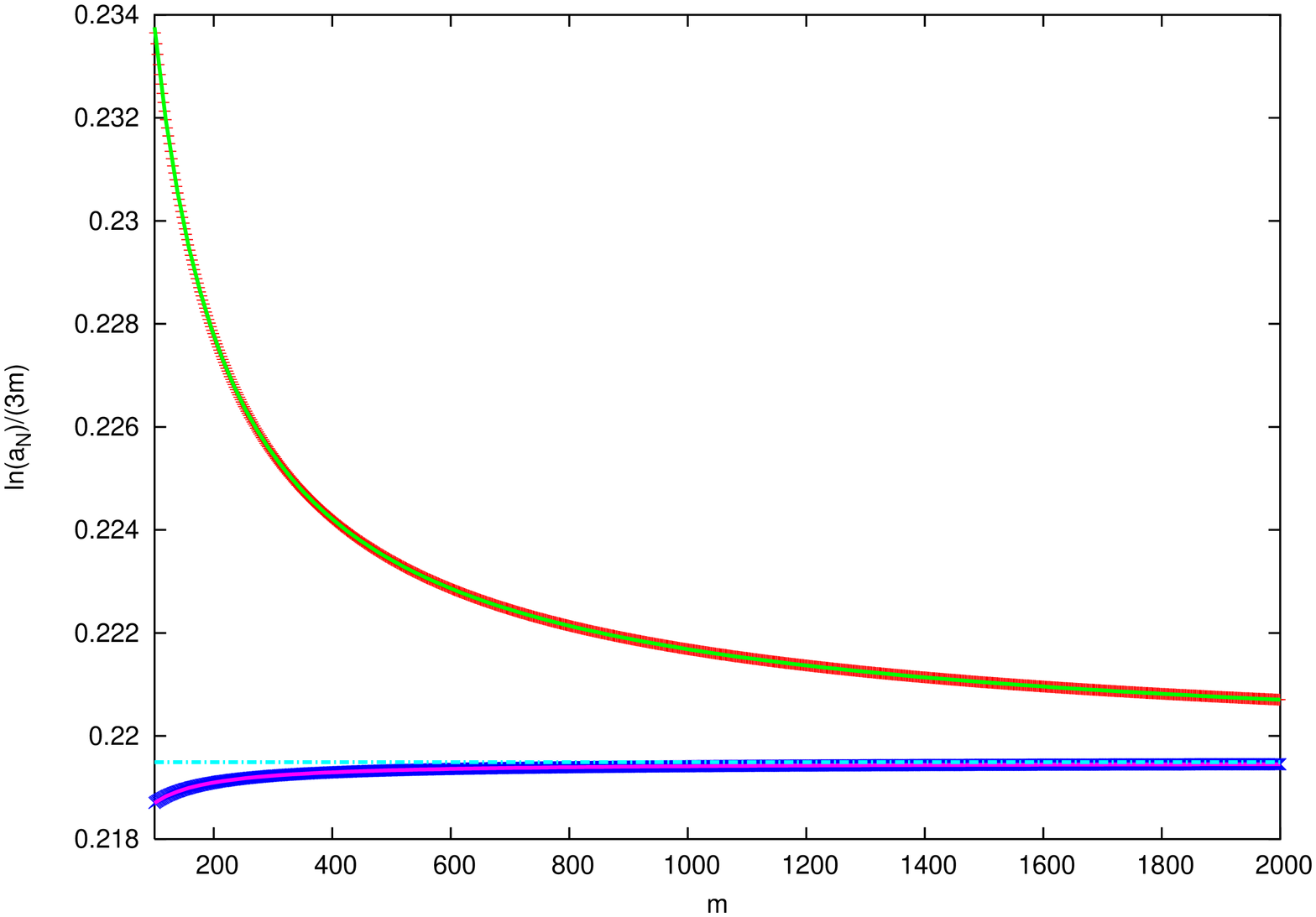}
  \end{minipage}\\
    \begin{minipage}[b]{\columnwidth}
      \centering 
      \includegraphics[angle=270,width=\columnwidth]{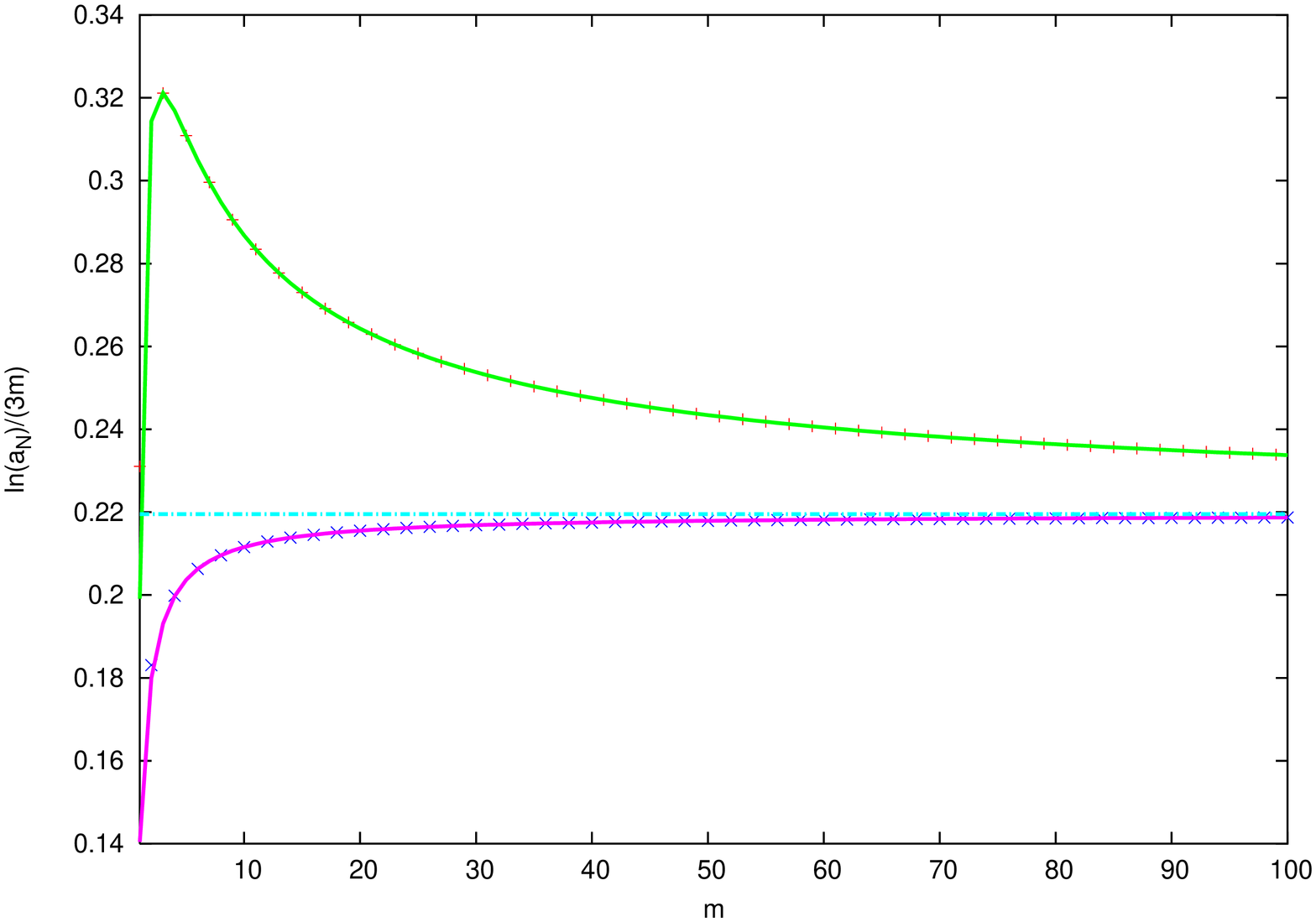}
  \end{minipage}%
  \caption{(Color online) 
    The original data of $\ln(a_N)/(mn)$ for $n=3$
    and the fitted curves.
    In each panel
    the data points and curve in the upper part
    are for odd $m$, and those in the lower part are
    for even $m$.
    The dashed horizontal line is $a_0^e(3) = 0.219493$
    from exact expression Eq. \ref{E:a_0_e}.
    The data $m \ge m_0 = 100$ are used in the fitting, and they are
    shown in the top panel. 
    In the bottom panel
    the same curves are shown together with the original data
    for $1 \le m < m_0 = 100$, which are not used in the fitting.
    \label{F:fit-3}} 
\end{figure}

\begin{figure}
    \begin{minipage}[b]{\columnwidth}
      \centering 
      \includegraphics[angle=270,width=\columnwidth]{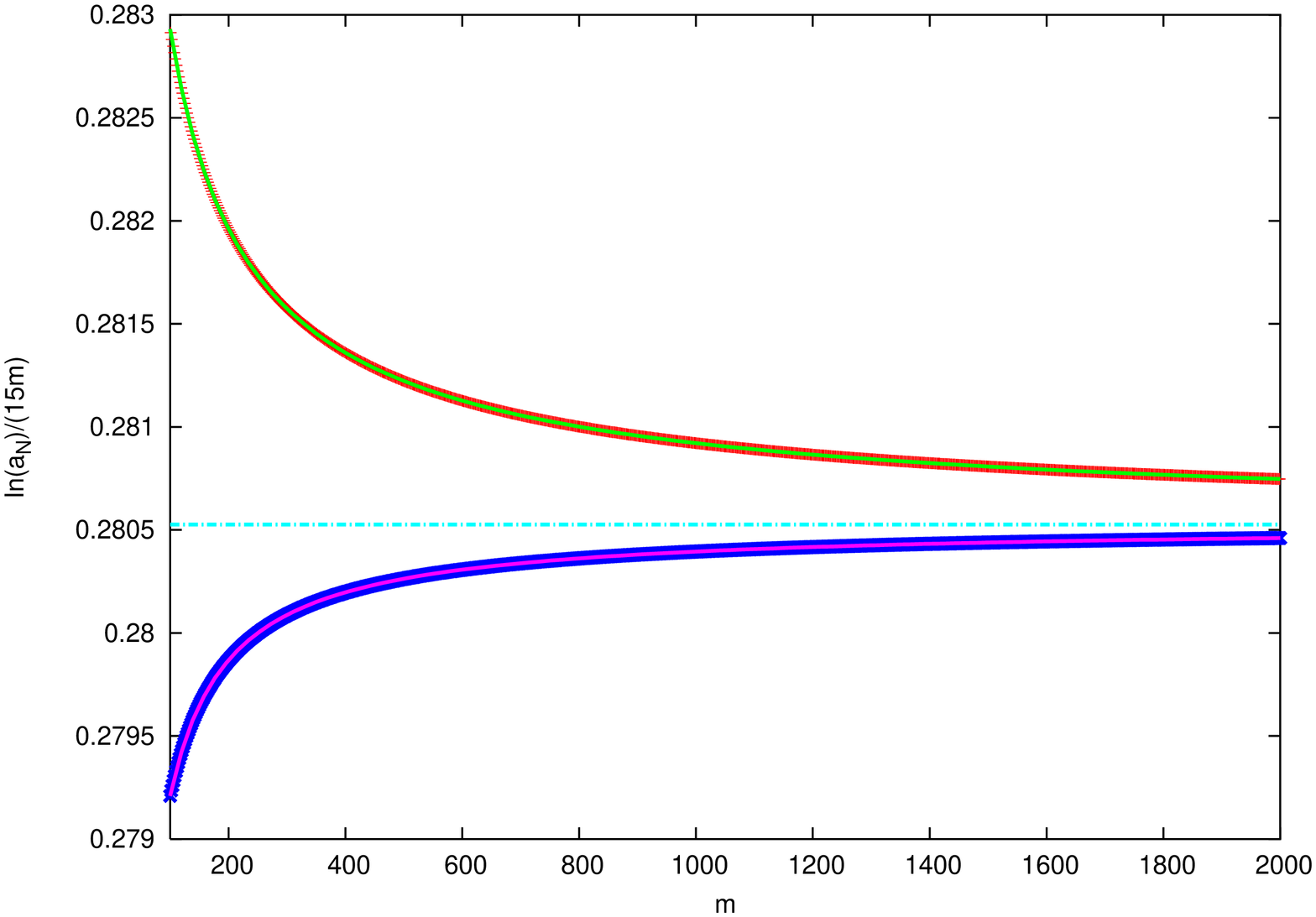}
  \end{minipage}\\
    \begin{minipage}[b]{\columnwidth}
      \centering 
      \includegraphics[angle=270,width=\columnwidth]{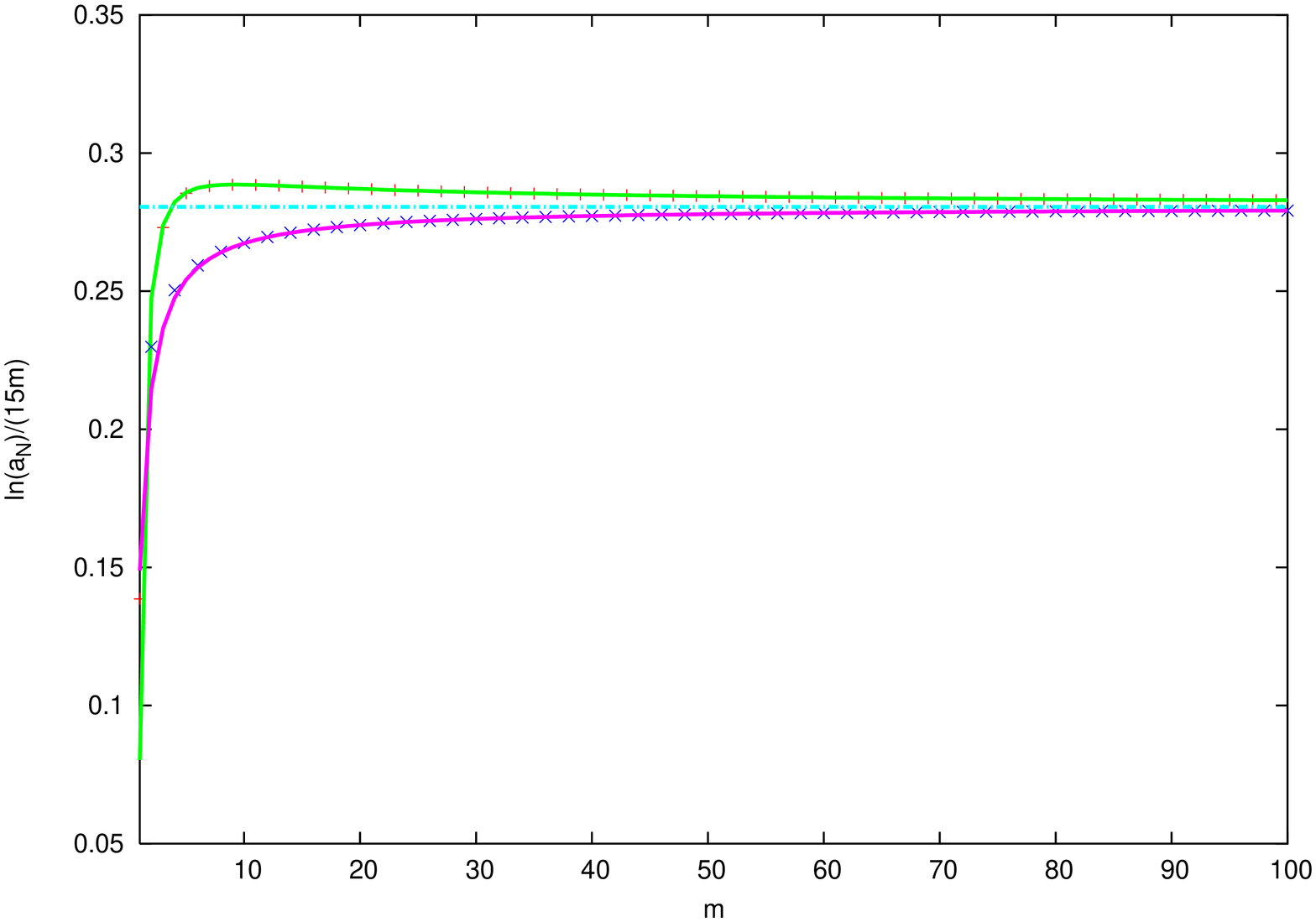}
  \end{minipage}%
  \caption{(Color online) The original data of $\ln(a_N)/(mn)$ for $n=15$
    and the fitted curves. 
    The dashed horizontal line is $a_0^e(15) = 0.280527$
    from exact expression Eq. \ref{E:a_0_e}.
    See the legend of Figure~\ref{F:fit-3}.
    \label{F:fit-15}} 
\end{figure}

In the following discussions, we use $n$ as the fixed width
of the strip lattice, and $m$ as the expanding length.
For each $n = 3, \dots, 19$, we calculate $a_N$ from $m=1$ to $m=2000$, 
except for $n=17$, where $m \le 1000$, and for $n=19$, where $m \le 500$.
Odd and even values of the same strip lattice are fitted separately.
The results of the fitting are shown in Table ~\ref{T:fit_odd}
for odd values of $m$
and Table ~\ref{T:fit_even} for even values of $m$.
In these fittings, only $m \ge m_0 = 100$ are used in the fitting.
The effect of $m_0$ on the fitting is discussed later.
The curves of fitting for two values of $n$, $n=3$ and $n=15$,
are shown in Figure~\ref{F:fit-3} and Figure~\ref{F:fit-15}, respectively.

Several features can be observed directly from these results
of fitting the data to Eq. \ref{E:fit}.
Firstly, for odd values of $m$, 
the coefficient $h$ of $(mn)^{-1} \ln(m+1)$
is equal to $1$, for \emph{all}
values of $n$ (Table \ref{T:fit_odd}).
This logarithm term is clearly absent when $m$ is even 
(Table \ref{T:fit_even}).
It is noted that there is also a logarithm term in the 
finite size correction of 
lattices where the single
vacancy is restricted at specific sites on the boundary of the
lattices \cite{Tzeng2003}. However, the logarithm correction there
is quite different from the logarithm correction discovered here.
Not only the sign and magnitude are different ($-1/2$ in \cite{Tzeng2003}),
more importantly, the logarithm correction in \cite{Tzeng2003}
only becomes significant when \emph{both} $m$ and $n$ become large.
On the contrary, the logarithm correction term for odd-by-odd lattices
appears in \emph{all} lattices, even when $n=1$ as shown in Eq. \ref{E:n1}.

Secondly, the linear term $a_0$ agrees exactly between odd $m$ values
and even $m$ values, and they both agree with 
the exact expression for \emph{infinitely} long strips of finite width $n$
\cite{Kasteleyn1961}:
\begin{equation} \label{E:a_0_e}
  a_0^e(n) = \frac{1}{n} \ln \left[ \prod_{l=1}^{\frac{n}{2}}
  \left( \cos \frac{l \pi}{n+1}
  + \left(1 + \cos^2 \frac{l \pi}{n+1} \right)^{\frac{1}{2}} 
  \right) \right].
\end{equation}
This agreement is expected, as $m$ goes into infinity,
the property of odd and even lattices should approach the same value.
If we fix the values of $a_0$ to those given by Eq. \ref{E:a_0_e},
and fit the data for the other parameters,
there is little changes in the fitting results.
The fitting results with fixed $a_0$ are shown in Tables \ref{T:fit_odd_fix}
and \ref{T:fit_even_fix}. For odd $m$, there is virtually no change in
$a_1$, $a_2$ and $h$, and for even $m$, there is no change in $a_1$.

From Figures~\ref{F:fit-3} and \ref{F:fit-15} we can see that there are
good matches between 
the fitted curves and
the original data, for both odd and even values of $m$. 
In both figures,
the top panels display the fitted curves and the data used in the fitting
($m \ge m_0  = 100$).
The bottom panels display the fitted curves and the data \emph{not}
used in the fitting
($1 \le m < m_0  = 100$).
Even for those data points that are not used in the fitting,
the match between fitted curves and the original data is good.
From these figures it can also be seen that the free energy
converges slower when $m$ is odd than when $m$ is even.

The effects of $m_0$ on the fitting results
are shown in Table \ref{T:fit-15} for $m \times 15$ lattices. 
From the table we can see that the fitting results converge quickly
as a function of $m_0$, the minimal value of $m$ used in the fitting.

\begingroup
\squeezetable
\begin{table*}
\caption{Fitting $\ln a_N/(mn)$ to Eq. \ref{E:fit} for \emph{odd} 
values of $m$.
Only data with $m \ge m_0 = 101$ are used in the fitting.
\label{T:fit_odd}}
\begin{ruledtabular}
\begin{tabular}{rdddddd}
 $n$  & \mbox{$a_0$} & \mbox{$a_1$} & \mbox{$a_2$} & \mbox{$a_3$} 
  & \mbox{$a_4$}
  & \mbox{$h$} \\\hline
 3  & 0.219493 
  & -0.11012 & -0.172442 
  & 0.127814 & -0.0964287 & 1\\ 
 5  & 0.252922 
  & -0.0562601 & -0.162238 
  & 0.0964196  & -0.0648308 & 1\\ 
 7  & 0.265557 
  & -0.0491038 & -0.152149 
  & 0.0711154  & -0.0466402   & 1\\ 
 9  & 0.272073 
  & -0.0514573 & -0.145208 
  & 0.0503154  & -0.0373426   & 1\\ 
 11 & 0.276016 
  & -0.0561077 & -0.140372 
  & 0.0319921  & -0.034609 & 1\\ 
 13 & 0.278649 
  & -0.0611249 & -0.136868 
  & 0.0151006  & -0.037443 & 1\\ 
 15 & 0.280527 
  & -0.0659264 & -0.134236 
  & -0.00090841 & -0.045399 & 1\\ 
 17 & 0.281932 
  & -0.0703454 & -0.132197 
  & -0.0163402 & -0.0583437 & 1\\ 
 19 & 0.283023 
  & -0.0743568 & -0.130576 
  & -0.0313702 & -0.076415 & 1
\end{tabular}
\end{ruledtabular}
\end{table*}
\endgroup

\begingroup
\squeezetable
\begin{table*}
\caption{Fitting $\ln a_N/(mn)$ to Eq. \ref{E:fit} for \emph{even}
 values of $m$.
Only data with $m \ge m_0 = 100$ are used in the fitting.
\label{T:fit_even}}
\begin{ruledtabular}
\begin{tabular}{rdddddd}
 $n$  & \mbox{$a_0$} & \mbox{$a_1$} & \mbox{$a_2$} & \mbox{$a_3$} 
  & \mbox{$a_4$}
  & \mbox{$h$} \\\hline
 3  & 0.219493 
 & -0.0791336 
 & -1.51574$e-08$
 & 1.46224$e-06$
 & -5.93864$e-05$
 & -8.41131$e-11$\\
 5  & 0.252922 
 & -0.102613  
 & -1.09579$e-08$ 
 & 4.6246$e-07$  
 & 1.1226$e-06$
 & -1.65381$e-10$\\
 7  & 0.265557 
 & -0.114309  
 & 4.47355$e-08$
 & -4.20078$e-06$
 & 0.000166702
 & 5.96032$e-10$\\
 9  & 0.272073 
 & -0.121309  
 & 4.4861$e-08$
 & -4.25669$e-06$
 & 0.000170225 
 & 7.58016$e-10$\\
 11 & 0.276016 
 & -0.125966  
 & 1.7632$e-09$ 
 & -6.40388$e-07$ 
 & 4.02706$e-05$
 & -9.03742$e-11$\\
 13 & 0.278649 
 & -0.129288  
 & -5.39512$e-09$
 & -9.39239$e-08$
 & 2.35758$e-05$
 & -3.1035$e-10$\\
 15 & 0.280527 
 & -0.131777  
 & 7.97321$e-08$
 & -7.59505$e-06$
 & 0.000305046
 & 2.24049$e-09$\\
 17 & 0.281932 
 & -0.13371   
 & 1.08088$e-07$
 & -8.71801$e-06$
 & 0.000312377
 & 4.42329$e-09$\\
 19 & 0.283023 
 & -0.135256  
 & 3.1797$e-07$
 & -2.0882$e-05$
 & 0.000640364
 & 1.91347$e-08$
\end{tabular}
\end{ruledtabular}
\end{table*}
\endgroup

\begingroup
\squeezetable
\begin{table*}
\caption{Fitting $\ln a_N/(mn)$ to Eq. \ref{E:fit} for \emph{odd}
  values of $m$,
  with $a_0$ fixed as $a_0^e$ given by Eq. \ref{E:a_0_e}.
  Only data with $m \ge m_0 = 101$ are used in the fitting.
\label{T:fit_odd_fix}}
\begin{ruledtabular}
\begin{tabular}{rd|ddddd}
 $n$  & \mbox{$a_0^e$} & \mbox{$a_1$} & \mbox{$a_2$} & \mbox{$a_3$} 
  & \mbox{$a_4$}
  & \mbox{$h$} \\\hline
 3  & 0.219493 
  & -0.11012 & -0.172442 & 0.12782 & -0.096693 & 1\\ 
 5  & 0.252922 
  & -0.0562601 & -0.162238 & 0.0964239 & -0.0649986 & 1\\ 
 7  & 0.265557 
  & -0.0491038 & -0.152149 & 0.0711184 & -0.04676   & 1\\ 
 9  & 0.272073 
  & -0.0514573 & -0.145208 & 0.0503178 & -0.0374353 & 1\\ 
 11 & 0.276016 
  & -0.0561077 & -0.140372 & 0.0319939 & -0.0346786 & 1\\ 
 13 & 0.278649 
  & -0.0611249 & -0.136868 & 0.0151016 & -0.0374833 & 1\\ 
 15 & 0.280527 
  & -0.0659264 & -0.134236 & -0.00090856 & -0.0453931 & 1\\ 
 17 & 0.281932 
  & -0.0703454 & -0.132197 & -0.0163431 & -0.0582417 & 1\\ 
 19 & 0.283023 
  & -0.0743568 & -0.130576 & -0.0313809 & -0.0760878 & 1
\end{tabular}
\end{ruledtabular}
\end{table*}
\endgroup

\begingroup
\squeezetable
\begin{table*}
\caption{Fitting $\ln a_N/(mn)$ to Eq. \ref{E:fit} for \emph{even}
  values of $m$,
  with $a_0$ fixed as $a_0^e$ given by Eq. \ref{E:a_0_e}.
  Only data with $m \ge m_0 = 100$ are used in the fitting.
\label{T:fit_even_fix}}
\begin{ruledtabular}
\begin{tabular}{rd|ddddd}
 $n$  & \mbox{$a_0^e$} & \mbox{$a_1$} & \mbox{$a_2$} & \mbox{$a_3$} 
  & \mbox{$a_4$}
  & \mbox{$h$} \\\hline
 3  & 0.219493 
 & -0.0791336 
 & 7.47107$e-09$ & -9.44524$e-07$ & 4.45608$e-05$ & 2.27359$e-11$\\
 5  & 0.252922 
 & -0.102613  & 1.94042$e-08$ & -2.4284$e-06$  & 0.000113707 & 9.98152$e-11$\\
 7  & 0.265557 
 & -0.114309  
 & 1.52227$e-08$ & -1.92429$e-06$ & 9.08621$e-05$ & 1.08499$e-10$\\
 9  & 0.272073 
 & -0.121309  
 & 1.33789$e-08$ & -1.63273$e-06$ & 7.50474$e-05$ & 1.28054$e-10$\\
 11 & 0.276016 
 & -0.125966  
 & 1.13578$e-08$ & -1.42994$e-06$ & 6.72433$e-05$ & 1.27535$e-10$\\
 13 & 0.278649 
 & -0.129288  
 & 1.87165$e-08$ & -2.35995$e-06$ & 0.000111213 & 2.48496$e-10$\\
 15 & 0.280527 
 & -0.131777  
 & 1.53596$e-08$ & -1.93301$e-06$ & 9.09871$e-05$ & 2.3596$e-10$\\
 17 & 0.281932 
 & -0.13371   
 & 2.90394$e-08$ & -3.27894$e-06$ & 0.000144225 & 6.13228$e-10$\\
 19 & 0.283023 
 & -0.135256  
 & 5.2761$e-08$ & -4.99251$e-06$  & 0.000194123 & 1.61671$e-09$
\end{tabular}
\end{ruledtabular}
\end{table*}
\endgroup

\begingroup
\squeezetable
\begin{table*}
\caption{The fitting of Eq. \ref{E:fit} for $n=15$
with different values of $m_0$, the minimal value of $m$
used in the fitting.
\label{T:fit-15}}
\begin{ruledtabular}
\begin{tabular}{rddddd}
 $m_0$  & \mbox{$a_1$} & 
          \mbox{$a_2$} & 
	  \mbox{$a_3$} & 
	  \mbox{$a_4$} & \mbox{$h$}\\\hline
1     &   -0.0646792  &  -0.141833   &  -0.0446463   & 0.0632163   & 0.996424\\
31    &   -0.0659261  &  -0.134254   &  0.000154746  & -0.0674233  & 1\\
51    &   -0.0659264  &  -0.134236   &  -0.000905825 & -0.0454885  & 1\\
101   &   -0.0659264  &  -0.134236   &  -0.000908559 & -0.0453931  & 1\\
201   &   -0.0659264  &  -0.134236   &  -0.000908559 & -0.0453931  & 1\\
301   &   -0.0659264  &  -0.134236   &  -0.000908559 & -0.0453931  & 1\\
\end{tabular}
\end{ruledtabular}
\end{table*}
\endgroup

From these fitting experiments it is clear that there is an
additional logarithm
term for the free energy of odd-by-odd lattices.
In this case, the asymptotic expression of $\ln a_N$ is given by
\begin{equation} \label{E:asymb}
 \ln a_N \approx  mn f_b + (m+n) f_s + \ln(m+1) + \ln(n+1)  
\end{equation}
where $f_b$ and $f_s$ are the ``bulk'' and ``surface'' terms,
respectively \cite{Fisher1961}:
\[
 f_b = G/\pi = 0.291560904
\]
and
\[
 f_s = G/\pi - \frac{1}{2} \ln(1+\sqrt{2}) = -0.14912589.
\]

With this asymptotic expression, we revisit the free energy $\ln a_N/n^2$
of \emph{odd} square lattices with size $n \times n$
shown in Figure~\ref{F:a_N} and investigate the origin
of the non-monotonicity and minimum found in $\ln a_N/n^2$ for odd lattices. 
For even square lattices, the exact results of close-packed
lattices give
the asymptotic expression of even square lattices as \cite{Ferdinand1967}
\begin{align*}
 \frac{\ln a_N}{n^2} &\approx 
 f_b 
 + \frac{2 f_b - \ln(1+\sqrt{2})}{n} 
 + \frac{\ln 2 + f_b - \ln(1+\sqrt{2})}{n^2} \\
 &= 0.291560904 - 0.298251779 n^{-1} + 0.1033344977 n^{-2},
\end{align*}
which increases monotonically with $n$ when $n \ge 1$.
In contrast, the
logarithm term in odd lattices 
introduces a minimum into the function.

\begin{figure}
  \centering
  \includegraphics[angle=270,width=\columnwidth]{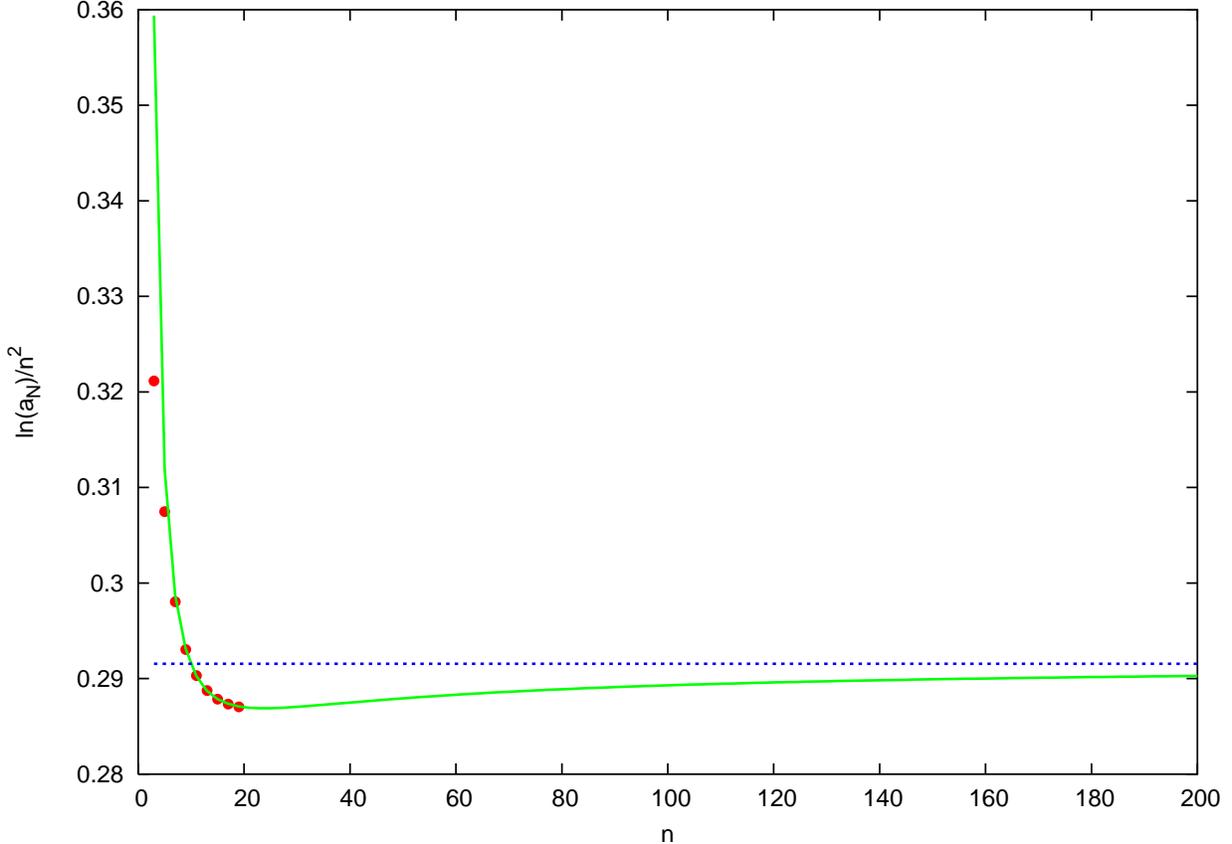}
\caption{(Color online) 
Fitting of $\ln a_N/n^2$ on odd square lattices to Eq. \ref{E:fit-square}.
Data in the range of $9 \le n \le 19$ are used in fitting.
The dashed horizontal line is the value in the thermodynamic limit:
$0.291560904$.
\label{F:a_N_fit}}
\end{figure}

For odd square lattices,
if we fit the data of $\ln a_N/n^2$ to the following equation, 
\begin{equation} \label{E:fit-square}
 \frac{\ln a_N}{n^2} = f_b + \frac{2 f_s}{n} 
 + \frac{b_2}{n^2} +  \frac{b_3}{n^3}
 +  \frac{2 \ln(n+1)}{n^2}
\end{equation}
we obtain $b_2 = -2.06412$
and $b_3 = 2.39028$.
The fitting uses data in the range of $9 \le n \le 19$.
The fitted curve together with the original data is shown
in Figure~\ref{F:a_N_fit}.
If we fit the data 
with the coefficient of $\ln(n+1)/n^2$ as a free parameter,
we would get $1.74647$ as the coefficient.
The derivation from the expected value of $2$
is attributed to the fact that the values of $n$ used in the fitting
are not large enough.
For the same reason, the values of $b_2$ and $b_3$ mentioned
above are of limited accuracy.
%
%

If we fit the same Eq. \ref{E:fit-square} to $\ln(n+1)/n^2$ 
of \emph{even} lattices with the coefficient of $\ln(n+1) / n^{2}$
as a free parameter (using data in the range of $100 \le n \le 1000$), 
we obtain 
$\ln a_N/n^2 \approx 0.291560904 - 0.298251779 n^{-1} 
+ 0.103144 n^{-2} + 0.00668442 n^{-3} 
+ 0.00003 \ln(n+1) / n^{2}$.
This result again shows the absence of the logarithm term
for the even lattices, as shown before in Table \ref{T:fit_even}
and Table \ref{T:fit_even_fix}.
The coefficient of $n^{-2}$ agrees well with the exact result.
The different behaviors of free energy
in odd and even square lattices are dominantly determined
by the presence of the logarithm term in odd square lattices.

%
%
%

%
%

\section{Discussion} \label{S:discussion}

We report here the exact values of $a_N$,
the coefficients of the leading term in the partition 
functions of dimers on  \emph{odd-by-odd} lattices 
through 
extensive computations using an extended method 
originally developed in Ref. \cite{Kong1999}.
At first we investigate $a_N$ on 
$(2k+1) \times (2k+1)$ odd square lattices, 
and from these exact values, we investigate 
the different 
behaviors of the average free energy of the odd lattices 
compared with that of the 
even lattices.
Although no exact solution 
exists for the odd lattices, the values of $a_N$ show some 
remarkable features (Table \ref{T:a_N}). 
Based on their number-theoretical properties,
we put forth several \emph{conjectures} about $a_N$.

The average free energy of the odd lattices approaches the
thermodynamic limit in a way that is quite different from
that of the even lattices, as well as that of the lattices where
a single vacancy is restricted at certain specific sites on the lattice
boundary (Figure~\ref{F:a_N}). 
Comparison with this latter case shows that the removal
of the restriction on the location of the vacancy 
changes the nature of the problem: while in both cases
there is a single vacancy in the whole lattice,
the restriction of the vacancy on the lattice boundary
significantly reduces the dimer configuration numbers.
In fact, when the vacancy is restricted on the  lattice boundary,
$a_N$ on a lattice of size $(2k+1) \times (2k+1)$ is asymptotically a 
$\sqrt{2k+1}$ multiplicative factor smaller than $a_N$
in an even-by-even lattice of size $(2k+2) \times 2k$ \cite{Tzeng2003}. 
On the contrary, the $a_N$ on the odd lattices discussed in this
report is \emph{bigger} than that of even-by-even lattices.
For example, 
the value $a_N$ of an odd-by-odd $19 \times 19$ lattice reported in
Table \ref{T:a_N} is $68$ times bigger than the value of $a_N$ 
for an even-by-even $18 \times 20$ lattice,
which is $a_N = 14766712169803333833186604776310955189771941$.

In Section \ref{S:FSC} we investigate $a_N$ on odd-by-odd strip lattices
of size $m \times n$, where $n$ is the fixed width of the lattice.
We find \emph{unambiguously} that there is a logarithm term
in the finite size correction
of the free energy $\ln a_N/mn$.  
This term is clearly absent in the free energy of even-by-even lattices.
We also demonstrate that 
it is this logarithm correction term that creates the distinct pattern
of free energy in odd lattices.
From these results we show that
in the asymptotic expression of the free energy per site 
for odd-by-odd lattices, 
in addition to the usual ``bulk'' and ``surface'' terms of
close-packed lattices
\cite{Fisher1961,Ferdinand1967},
we have an additional term $\ln(m+1)(n+1)$ as shown in Eq. \ref{E:asymb}.
It is also noted that this logarithm finite size correction term
is present for strip lattices of any width $n \ge 1$.
This is in contrast to the logarithm term in
lattices where the vacancy is restricted to the lattice boundary,
where the term only becomes significant
when \emph{both} $m$ and $n$ are large \cite{Tzeng2003}.

Since the exact solution to the close-packed dimer model was
discovered in 1961,
little progress has been made for the general
two-dimensional monomer-dimer problem. 
The problem is usually considered to be 
computationally intractable \cite{Jerrum1987}.
More precisely, in the language of computational complexity,
it has been shown to be in the \emph{``\#P-complete''} class.
\#P-complete class
plays the same role for counting problems (such as counting dimer
configurations in two-dimensional lattices,
as discussed in this report)
as the more familiar \emph{NP-complete} class plays for the
decision problems.
\#P-complete problems belong to the class of problems
called \#P class, which has the same status as the \emph{NP} class 
for the decision problems.
Among the problems in \#P class,
\#P-complete problems are the ``hardest'': every problem in \#P class
can be \emph{reduced} to them in polynomial-time. 
Hence if \emph{any} problem in the \#P-complete class 
is found to be solvable, every problem in \#P class
is solvable.
Currently it is not clear whether there exists any such solution
to the \#P-complete class problems, and ``P verse NP'' problem
is the perhaps the major
outstanding problem in theoretical computer science.

Although currently we still do not have an exact solution to the problem of
packing dimers on odd-by-odd lattices,
the number-theoretical properties demonstrated for the square lattices
and the unambiguous  logarithm term in the
finite size correction point to the possibility that the model
may actually be solvable.
It is hoped 
that the conjectures and the exact coefficients of 
the partition functions reported here would give 
some hints to the elusive exact solutions, 
and act as references for other approaches to 
the problem, such as Monte Carlo simulations.  
It would be interesting to see if other 
unsolved models in statistical mechanics show 
the similar patterns in their 
enumerations.  It is our hope that the results shown here 
will open up new avenues
and stimulate new 
mathematical and computational approaches to the unsolved statistical 
models.

\bibliography{md}

\end{document}